\documentclass[journal]{IEEEtran}

\usepackage{cite}
\usepackage{graphicx}
\usepackage[cmex10]{amsmath}
\usepackage{bbm}
\usepackage{amsfonts,amssymb,amsbsy}
\usepackage{algorithmicx}
\usepackage{algpseudocode}
\usepackage{algorithm}
\usepackage{dsfont}
\usepackage{color}
\usepackage{verbatim}

\graphicspath{{fig/}}

\ifCLASSOPTIONcompsoc
  \usepackage[caption=false,font=normalsize,labelfont=sf,textfont=sf]{subfig}
\else
  \usepackage[caption=false,font=footnotesize]{subfig}
\fi

\algnewcommand\algorithmicinput{\textbf{INPUT:}}
\algnewcommand\INPUT{\item[\algorithmicinput]}

\algnewcommand\algorithmicoutput{\textbf{OUTPUT:}}
\algnewcommand\OUTPUT{\item[\algorithmicoutput]}

\algnewcommand\algorithmicinit{\textbf{Initialization:}}
\algnewcommand\Init{\item[\algorithmicinit]}

\algdef{SE}[DOWHILE]{Do}{doWhile}{\algorithmicdo}[1]{\algorithmicwhile\ #1}%

\DeclareMathOperator*{\minimize}{minimize}
\DeclareMathOperator{\subjectto}{subject~to}

\makeatletter
\newenvironment{varsubequations}[1]
 {%
  \addtocounter{equation}{-1}%
  \begin{subequations}
  \renewcommand{\theparentequation}{#1}%
  \def\@currentlabel{#1}%
 }
 {%
  \end{subequations}
 }
\makeatother

\newcommand{\bt}{\boldsymbol{t}}
\newcommand{\br}{\boldsymbol{r}}
\newcommand{\bs}{\boldsymbol{s}}
\newcommand{\bw}{\boldsymbol{w}}
\newcommand{\bx}{\boldsymbol{x}}
\newcommand{\by}{\boldsymbol{y}}
\newcommand{\bz}{\boldsymbol{z}}
\newcommand{\bv}{\boldsymbol{v}}
\newcommand{\blambda}{\boldsymbol{\lambda}}
\newcommand{\NN}{\mathcal{N}}
\newcommand{\KK}{\mathcal{K}}

\newtheorem{theorem}{Theorem}

\newlength{\mywidth}
\makeatletter
\newif\ifhbonecolumn
\@ifclasswith{IEEEtran}{onecolumn}{\hbonecolumntrue}{\hbonecolumnfalse}
\makeatother
\begin{document}

\title{Energy-Efficient Cell Activation,
User Association, and Spectrum Allocation
in Heterogeneous Networks%
\thanks{
B.~Zhuang was with the Department of Electrical
  Engineering and Computer Science at Northwestern University,
  Evanston, IL, 60208, USA.
  He is now with Samsung, Inc., San Diego, CA, USA.
D.~Guo and M.~L.~Honig are with the Department of Electrical
  Engineering and Computer Science at Northwestern University,
  Evanston, IL, 60208, USA.  
}%
\thanks{This work was supported in part by a gift from Futurewei Technologies
  and by the National Science Foundation under Grant No.~CCF-1231828.}}

\author{
Binnan Zhuang,
Dongning Guo,
and Michael L. Honig}

\maketitle

\begin{abstract}
Next generation (5G) cellular networks are expected to be supported by
{an} extensive infrastructure
with many-fold increase in the number of cells per
{unit area compared to today}.
{The total} energy consumption of base transceiver stations (BTSs)
{is an important issue}
for both economic and environmental reasons. In this paper, an optimization-based framework is proposed for energy-efficient {global} radio resource management in heterogeneous wireless networks.
Specifically,
{with stochastic arrivals of known rates intended for users}, the smallest set of BTSs is activated with jointly optimized user association and spectrum allocation to stabilize
the network first {and then minimize the delay}. The scheme can be carried out periodically on a relatively slow timescale to adapt to aggregate traffic variations and average channel conditions.
Numerical results show that the proposed scheme significantly reduces the energy consumption and increases the quality of service compared to existing schemes in the literature.
\end{abstract}

\section{Introduction}
\label{sec:Intro}

Commercial wireless networks are evolving towards higher frequency reuse by deploying smaller cells to meet increasing demand for mobile data services.  In a heterogeneous network (HetNet), macro cells provide for wide area coverage and for serving highly mobile users, whereas dense deployment of femto, pico, and/or micro cells, possibly with distributed antennas, can support much higher data rates per unit area.

The energy consumption 
due to information and communication technologies worldwide is rising 
rapidly~\cite{
fehske2011global 
}.
As the number of base transceiver stations (BTSs) increases,
it is ever more important to manage their energy consumption.
{An active macro BTS consumes 40 to 80 watts on transmission,
whereas the total power consumption is typically hundreds to well
over 1,000 watts, which includes the power for signal processing,
computation, cooling, and radio frequency power amplification
(see, e.g., \cite{
CorZel10MCOM,
li2011energy-efficient, 
oh2011toward, 
mclaughlin2011techniques, 
hu2014energy}
~and references therein).}
{Hence, savings from transmit power control alone are relatively limited.
Much more significant power savings can be accomplished by}
turning a BTS off or 
switching it to deep sleep mode.

{In this paper, we study how to support given traffic with as few
active small cells as possible to conserve energy in a HetNet.}
{Because} it may take many seconds
to reactivate 
{a BTS in deep sleep}~\cite{FreMob11VTC}, 
the on-off decision
{should rely} on the {\em aggregate} traffic and average channels conditions collected over a period, possibly lasting a minute or more.
Cell activation/deactivation thus occurs on a much slower timescale than channel-aware scheduling, which typically occurs over time slots of a few milliseconds~\cite{AstDah09MCOM}.
The central problem considered here is how to jointly optimize
spectrum allocation and user association 
to minimize the number of active cells.
Without loss of generality, we focus on
downlink data transmissions.

The problem formulation in this paper builds upon
prior work~\cite{
zhuang2015traffic-driven}.
The following two features distinguish this paper and~\cite{
zhuang2015traffic-driven}
from most existing work on
energy-efficient resource management in the literature, including~\cite{
MarChi09ICCW,
ZhoGon09MobiCom,
son2011base,
pollakis2012base,
niu2012energy-aware,
oh2013dynamic,
soh2013energy,
yang2013energy,
hossain2014traffic-aware,
li2014energy-efficient,
lim2014energy-efficient
}.
First, we consider stochastic packet arrivals to user traffic queues
in lieu of static data rate requirements.
{Stochastic traffic better models the challenges for small cells as they
have much more pronounced traffic variations than macro cells.}
The proposed slow timescale 
resource management adapts to the average traffic load, thus avoiding
frequent 
on/off {updates due to varying user equipment (UE) data rates.}
{Second, the formulation here facilitates the maximum amount
of spectrum agility.  Specifically, by considering all possible reuse
patterns,}
arbitrary (possibly nonconsecutive) spectrum can be allocated
to a link from any BTS to any UE 
(see also~\cite{
KuaUtsDot2014arxiv,
dotzler2010fractional}).
This is in contrast to the
{full-spectrum reuse (i.e.,} all BTSs use all available spectrum{) 
assumed} in~\cite{
MarChi09ICCW,
ZhoGon09MobiCom,
son2011base,
pollakis2012base,
niu2012energy-aware,
oh2013dynamic,
soh2013energy,
yang2013energy,
hossain2014traffic-aware,
li2014energy-efficient
}.


The optimization problem formulated here 
is a mixed integer program,
{where cell activation decisions are expressed as binary variables.}
The problem is solved 
numerically using an iterative algorithm based on reweighted $\ell_1$ minimization~\cite{candes2008enhancing}.
The method was 
interpreted as 
majorization-minimization 
in~\cite{sriperumbudur2011majorization}.
Reference~\cite{pollakis2012base}
{uses the same method to solve the cell activation and user
association problem under full-spectrum reuse and static rate requirements.}

Numerical results show that the proposed 
resource management method achieves
significant energy savings
{as well as throughput gains in a typical HetNet.
In particular, the performance advantages of spectrum agility is
demonstrated by comparing with the method in~\cite{pollakis2012base}. As previously noted in~\cite{zhuang2015traffic-driven},}
spectrum agility is 
crucial for improving network throughput.
Intuitively, 
improved spectrum allocation
allows more BTSs to be turned off.
Furthermore,
a silent BTS causes no
interference to other cells, so that the remaining BTSs
may attain higher spectral efficiencies
~\cite{ZhuGuo2012Allerton}. This may present 
additional opportunities 
for turning off BTSs
and adding more energy savings.

The 
proposed method offers a different tradeoff between performance and
complexity than in~\cite{pollakis2012base}.
The scheme in~\cite{pollakis2012base} is computationally feasible for
a HetNet with hundreds of BTSs, whereas the 
method here is
feasible for a cluster of at most 20 to 30 BTSs due to a large
number of additional spectrum allocation variables in the optimization problem.

The remainder of this paper is organized as follows.
The system model is introduced in Section~\ref{sec:SysMod}.
The optimization problems
are presented in Section~\ref{sec:Opt}, and
the algorithms are introduced in Section~\ref{sec:Alg}.
Simulation results are presented in Section~\ref{sec:Sim},
and conclusions are drawn in Section~\ref{sec:Con}.

\section{System Model}
\label{sec:SysMod}

We consider the downlink of a HetNet with $n$ BTSs,
including $n_1$ macro BTSs and $n_2=n-n_1$ pico BTSs.
Denote the set of macro 
and pico BTSs as $\NN_1=\{1,\cdots,n_1\}$ and $\NN_2=\{n_1+1,\cdots,n\}$, respectively.
Let $\NN=\{1,\cdots,n\}$ denote the set of all the BTSs.
All BTSs operate on the same licensed band of $W$ Hz. The frequency resources are assumed to be homogenous on a slow timescale.

The key to total spectrum agility is the notion of
{\em pattern}~\cite{zhuang2015traffic-driven}.
A pattern
over a set of time-frequency resources is a subset of transmitters sharing the
resources. {In the downlink, a pattern $A$ is a subset of $\{1,\dots,n\}$,
and all BTSs in $A$}
have simultaneous access to the frequency band associated with the pattern.
Assuming known transmit power spectral densities (PSDs), a particular
pattern determines the signal-to-interference-plus-noise ratio (SINR)
and hence the spectral efficiencies of all links in the network.
The allocation problem can then be
formulated as how to divide
the resources among all $2^n$ patterns.
{In a simple example with only 2 BTSs, there are only 
patterns: $\{1\}$, $\{2\}$, $\{1,2\}$, and $\emptyset$, which denote the patterns used respectively by BTS 1 and BTS 2 exclusively, the pattern used by both BTSs, and the pattern used by none, respectively.  For every pattern $A\subset\NN$,} let $y_A$ be the fraction of total bandwidth allocated to {it}.
Clearly,
\begin{align}
  \sum_{A\subset \NN}y_A=1,
\end{align}
and any efficient allocation would
{set $y_{\emptyset}=0$}.


\begin{figure}[!t]
\includegraphics[width=\mywidth]{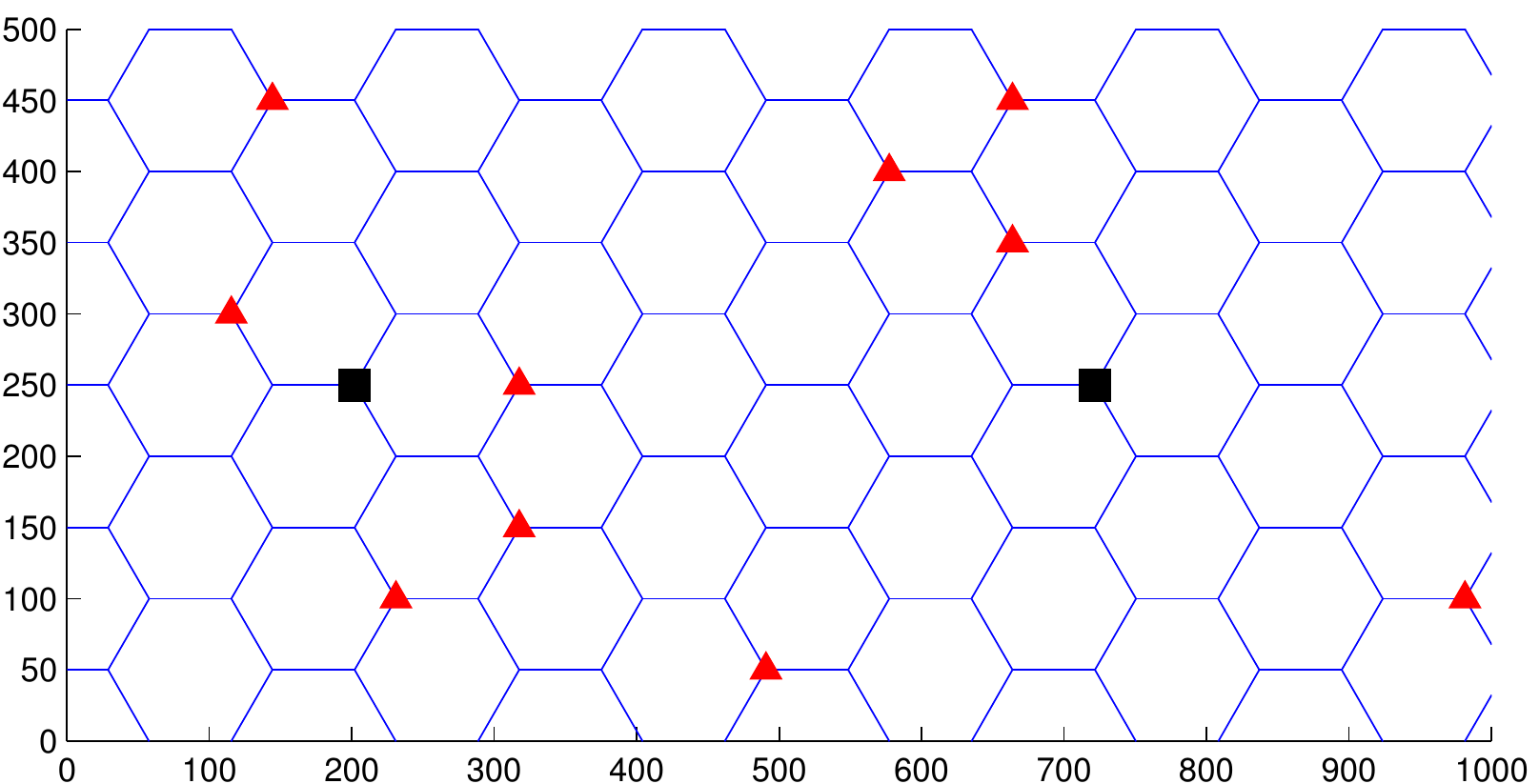}
\centering
\caption{The topology of a HetNet with $n_1=2$ macro BTSs, $n_2=10$ randomly deployed pico BTSs and $k=66$ user groups.}
\label{fig:topo}
\end{figure}

At the slow timescale considered it is reasonable to treat users near
each other with similar
quality of service (QoS) requirements as a {\em group}.
Denote the set of all user groups as $\KK=\{1,\cdots,k\}$.
An example HetNet model with $n=12$ BTSs and $k=66$ user groups is shown in Fig.~\ref{fig:topo}, where each group of UEs is assumed to be located at the center of each hexagon (as in~\cite{ZhuGuo2012Allerton}).
{For ease of characterizing the delay as the objective,}
the aggregate traffic of group $j$ UEs is modeled by Poisson traffic arrivals with arrival rate $\lambda_j$. The packet length is exponentially distributed with average length $L$.
{It is possible to adopt a different traffic and queueing model
(see, e.g.,~\cite{zhou2016allocation}).}


{We assume} each BTS assigns different spectrum resources to different user groups.
{This} can be viewed as statistically multiplexing the packet streams from different user groups. The packets from the same group of UEs are served according to a `first in first out' policy. Hence, each user group effectively has a virtual queue.
We {also} assume multiple BTSs can serve the same group of UEs.\footnote{On a slow timescale this can be realized by letting different BTSs serve different subsets of individual UEs.}

We assume BTS $i$ transmits to its UEs with fixed flat transmit PSD, $p_i$. The spectral efficiency of the link
{$i\to j$}
under pattern $A$ depends on the receive power and the interference. For concreteness in obtaining numerical results, we use Shannon's formula to obtain:
\begin{align}
\label{eq:SE}
s^{i\to j}_A=\frac{W\,\mathbbm{1}(i\in A)}{L}\log_2\left(1+\frac{p_ig^{i\to j}}{\sum_{i'\in A\setminus\{i\}}p_{i'}g^{i'\to j}+n_j}\right)
\end{align}\noindent
in packets/second,
where $\mathbbm{1}(i\in A)=1$ if $i\in A$ and $\mathbbm{1}(i\not\in A)=0$ otherwise, $g^{i\to j}$ is the power gain of the link
{$i\to j$}, and $n_j$ is the noise PSD at group $j$ UEs. The link gain $g^{i\to j}$ includes pathloss and shadowing effects over the slow timescale considered in this paper. Hence $g^{i\to j}$ is a constant in each decision period independent of the frequency. If small scale fading is included in $g^{i\to j}$ on the slow timescale, then ergodic spectral efficiency must be used instead of~\eqref{eq:SE}, since the decision period spans many
{coherence time intervals}.

All packets intended 
for group $j$ UEs arrive at an M/M/1 queue. 
Denote the bandwidth
allocated to BTS $i$ to serve group $j$ UEs under pattern $A$ as $x^{i\to j}_A$.
The service rate for this queue contributed by BTS $i$ under pattern $A$ is $s^{i\to j}_Ax^{i\to j}_A$.
{This rate is guaranteed regardless of the activities of the other BTSs and queues.}
The total service rate for queue $j$ is a linear function of the
bandwidths: 
\begin{align}
\label{eq:Rate}
r_j=\sum_{i\in \NN}\sum_{A\subset \NN} s^{i\to j}_Ax^{i\to j}_A\quad\text{packets/second}.
\end{align}\noindent
Hence the average packet sojourn time of the M/M/1 queue
for group $j$ UEs takes a simple form~\cite{Nel95Spinger}:
\begin{align}
\label{eq:IndDelay}
 t_j=\frac{1}{(r_j-\lambda_j)^+} \quad\text{seconds,}
\end{align}\noindent
where $(x)^+=\max\{0,x\}$, so $t_j=+\infty$ if $r_j\le\lambda_j$.

\section{{Optimization Framework}}
\label{sec:Opt}
\subsection{Baseline {Formulation}} 
\label{subsec:ProbStat}

The basic energy-efficient 
{resource} allocation problem is formulated as:
\begin{varsubequations}{P1}
\label{eq:Opt}
\begin{align}
{\minimize \limits_{\br,\bx,\by,\bz}} \quad& \sum_{i\in \NN_2}c_iz_i &\label{eq:OptObj}\\
\subjectto\;\; &z_i\in\{0,1\},
\quad \forall i\in \NN_2\label{eq:OptCon6}\\
&r_j=\sum_{i\in \NN}\sum_{A\subset \NN} s^{i\to j}_Ax^{i\to j}_A,
\quad \forall j\in \KK \label{eq:OptCon1}\\
&\sum_{j\in \KK}x^{i\to j}_A
  { \leq } y_A,
\quad \forall i\in \NN,\; A\subset \NN\label{eq:OptCon3}\\
&\sum_{A\subset \NN}\sum_{j\in \KK}x^{i\to j}_A\leq z_i,
\quad \forall i\in \NN_2\label{eq:OptCon7}\\
&x^{i\to j}_A 
\geq0,
\quad \forall i\in \NN,\; j\in \KK,\; A\subset \NN\label{eq:OptCon5}\\
&r_j-\lambda_j\geq \tau_j^{-1}, 
\quad \forall j\in \KK \label{eq:OptCon2}\\
&\sum_{A\subset \NN}y_A=1&\label{eq:OptCon4}
\end{align}
\end{varsubequations}
where:
\begin{itemize}
\item The variables $\br$, $\bx$, $\by$, {and} $\bz$ are the vector forms of
  $\left(r_j\right)_{j\in \KK}$,
  $\big(x^{i\to j}_A\big)_{i\in \NN,\;j\in \KK,\;A\subset \NN}$,
  $\left(y_A\right)_{A\subset \NN}$, and $\left(z_i\right)_{i\in \NN_2}$,
  respectively.
\item \eqref{eq:OptObj} is the total energy cost of the network, where
  $z_i$ constrained by~\eqref{eq:OptCon6} is a binary variable
  indicating whether pico BTS $i$ is on or off, and $c_i$ is the 
  power cost of pico BTS $i$ if it is active.
{While} the macro BTSs are {assumed to be} always on to provide
basic coverage, {it is easy to include their on-off decisions as
  variables as well.}
\item \eqref{eq:OptCon1} is the service rate of each user group as given in~\eqref{eq:Rate}.

\item \eqref{eq:OptCon3} guarantees
  {that, for every pattern $A$, the total bandwidth a BTS
    allocates to all groups does not exceed the bandwidth assigned to pattern $A$.}

\item \eqref{eq:OptCon7} states that a pico BTS $i$ 
{uses} no spectrum 
if it is off and uses at most 
{1} unit {of} bandwidth if it is on.

\item \eqref{eq:OptCon5} {constrains all bandwidths to be nonnegative.}

\item \eqref{eq:OptCon2} is the QoS constraint for each user
    group, where $\tau_j$ is the maximum delay {allowed}
    for group $j$ UEs.
    The constraint {is equivalent to letting $t_j$ in}%
    ~\eqref{eq:IndDelay} 
    {satisfy
    $0\leq t_j\leq\tau_j$}.
  {In lieu} of QoS, quality of experience (QoE) can be considered using the proposed framework as well. To achieve this, each user group can be further divided into smaller groups representing different rate, delay and other QoS preferences. Also, QoE may be modeled a general user-dependent utility function $Q_j(\lambda_j,r_j)$.

\item \eqref{eq:OptCon4} constrains the total system bandwidth to be one unit.
\end{itemize}

{The formulation~\ref{eq:Opt}
addresses} joint user association, spectrum allocation, and BTS
activation {all under one framework}.
The objective and all the constraints are linear.
Solving~\ref{eq:Opt} 
minimizes the energy cost by turning off
pico BTSs {not needed to satisfy} 
the delay requirements. 

\subsection{Structure of the Optimal Solution} 
\label{subsec:Prop}
One concern with~\ref{eq:Opt} is that the
spectrum is divided into up to $2^n$ patterns (segments), which may be
impractical for all but very small networks.  Fortunately,
we can use Carath\'eodory's Theorem to show that there exists an
optimal allocation that uses {no more than $k$} 
patterns.

\begin{theorem}\label{thm:spec}
  There exists an optimal solution of~\ref{eq:Opt} in which $\by$ is
  $k$-sparse, i.e.,
\begin{align}
\label{eq:Thm1}
\bigg|\{A~|~y_A>0,\; A\subset \NN \}\bigg|\leq k.
\end{align}
\end{theorem}

\begin{IEEEproof} 
  {We first reformulate~\ref{eq:Opt} by defining a new set of variables $\{v^{i\to j}_A\}$, which represent the fraction  of spectrum under pattern $A$ that BTS $i$ allocates to group $j$.
  The constraints~\eqref{eq:OptCon1}--\eqref{eq:OptCon5} then become:
\begin{varsubequations}{P2}
\label{eq:P2}
\begin{align}
&r_j=\sum_{A\subset \NN} \left(\sum_{i\in \NN} s^{i\to j}_A v^{i\to j}_A\right) y_A,
\quad \forall j\in \KK \label{eq:P2:r}\\
&\sum_{j\in \KK}v^{i\to j}_A 
  { \leq } 1,
\quad \forall i\in \NN,\; A\subset \NN\label{eq:P2:v1}\\
&\sum_{A\subset \NN}\sum_{j\in \KK}v^{i\to j}_A y_A \leq z_i,
\quad \forall i\in \NN_2\label{eq:P2:vyz}\\
&v^{i\to j}_A 
\geq0,
\quad \forall i\in \NN,\; j\in \KK,\; A\subset \NN. \label{eq:P2:v0}
\end{align}%
\end{varsubequations}%
This problem,~\ref{eq:P2}, where $\{v^{i\to j}_A\}$ replaces $\{x^{i\to j}_A\}$, is clearly equivalent to~\ref{eq:Opt}. Solving~\ref{eq:P2}, the actual
spectrum allocations are given by $x^{i\to j}_A = y_A v^{i\to j}_A$.
}


In the remainder of this proof, we show that if a solution $(\br,\bv,\by,\bz)$ to~\ref{eq:P2}
exists, then there exists a $k$-sparse $\by^*$ and rate tuple
$\br^*$ such that $(\br^*,\bv,\by^*,\bz)$ is feasible.  This attains the original objective~\eqref{eq:OptObj} because $\bz$ remains the same.  We shall verify the feasibility constraints pertaining to $(\br^*,\by^*)$,
including~\eqref{eq:OptCon2},~\eqref{eq:OptCon4},~\eqref{eq:P2:r},
and~\eqref{eq:P2:vyz}.

Suppose $\by$ has more than $k$ nonzero elements.  Let its
support be 
$\mathsf{S}$ ($y_A=0$ if $A\notin\mathsf{S}$).
Let us 
define
a $k$-vector $\bt_A$ for every $A\in\mathsf{S}$ with its elements
determined by $t_{j,A} = \sum_{i\in\NN} s^{i\to j}_A v^{i\to j}_A$.
  According to~\eqref{eq:P2:r}, a convex combination of the vectors
  $(\bt_A)_{A\in\mathcal{S}}$ with $(y_A)_{A\in\mathsf{S}}$ as
  coefficients form the optimal
  rate tuple: 
  $\br = \sum_{A\subset\NN} y_A \bt_A$
  By Carath\'eodory's Theorem, $\br$ can be represented as a
  convex combination of no more than $k+1$ of those vectors, denoted as $(t_A)_{A\in\mathsf{S}^*}$.
Moreover,
{either $\br$ is on the boundary of the convex hull of
$(\bt_A)_{A\in\mathsf{S}^*}$ or it is an interior point.
In either case, there exists $\br^*$ on the boundary that
dominates $\br$ in every dimension.
Clearly, $\br^*$ is the convex combination of at most $k$
vectors from $(\bt_A)_{A\in\mathsf{S}^*}$.  Therefore, there exists
$\by^*$ satisfying~\eqref{eq:OptCon4} with support
$\mathsf{S}^{**} \subset \mathsf{S}^* \subset \mathsf{S}$, such that
$|\mathsf{S}^{**}|\le k$ and
\begin{align}
  \br^* = \sum_{A\in\mathsf{S}^{**}} y^*_A \bt_A
\end{align}
which implies~\eqref{eq:P2:r}.
Because $\br^*$ dominates $\br$ in every dimension,~\eqref{eq:OptCon2}
is satisfied.
It remains to show that~\eqref{eq:P2:vyz} holds.
For every $i$ with $z_i=1$,~\eqref{eq:P2:vyz}
is satisfied due to~\eqref{eq:OptCon4} and~\eqref{eq:P2:v1}.
For every $i$ with $z_i=0$,~\eqref{eq:P2:vyz} requires that
$v^{i\to j}_A=0$ for every $j$ and every $A\in\mathsf{S}$ with $y_A>0$,
which implies that~\eqref{eq:P2:vyz} remains true for $\by^*$,
because its support is dominated by that of $\by$.}
This completes the proof.
\end{IEEEproof}

\subsection{Comparison with Full Spectrum Reuse~\cite{pollakis2012base}}
\label{subsec:Exist}
The algorithm to be introduced for solving~\ref{eq:Opt} is related to the majorization-minimization approach in~\cite{pollakis2012base}, which is based on reweighted $\ell_1$ minimization proposed in~\cite{candes2008enhancing}.
To facilitate a fair comparison, 
we formulate an analogous problem to the one in~\cite{pollakis2012base} for
cell activation and user association under the 
current framework.
With minor changes, 
the optimization problem in~\cite{pollakis2012base} becomes:
\begin{varsubequations}{P3}
\label{eq:Opt2}
\begin{align}
{\minimize \limits_{\br,\bx,\bz}} \quad& \sum_{i\in \NN_2}c_iz_i \quad\label{eq:Opt2Obj}\\
\subjectto\;\; &z_i\in\{0,1\},\quad\forall i\in \NN_2\label{eq:Opt2Con4}\\
&r_j=\sum_{i\in \NN} s^{i\to j}_{\NN} x^{i\to j},\;\quad\forall j\in \KK \label{eq:Opt2Con1}\\
&\sum_{j\in \KK}x^{i\to j}\leq z_i, \quad\forall i\in \NN\label{eq:Opt2Con5}\\
&x^{i\to j}\geq0, \quad\forall i\in \NN,\;\forall j\in \KK\label{eq:Opt2Con3}\\
&r_j-\lambda_j\geq \tau_j^{-1}, 
\quad\forall j\in \KK \label{eq:Opt2Con2}
\end{align}
\end{varsubequations}%
{where all variables have the same physical meaning as in~\ref{eq:Opt}.}
The {key} 
change from~\ref{eq:Opt} to~\ref{eq:Opt2} is that only 
full spectrum 
reuse 
is allowed, 
where the spectral efficiency of the link {$i\to j$}
{is} $s_{\NN}^{i\to j}$.  \ref{eq:Opt2} optimizes
{user} association
and the pico BTS on/off selection 
to minimize the energy cost.
The performance of allocations based on~\ref{eq:Opt} and~\ref{eq:Opt2} will be compared in Section~\ref{sec:Sim} using numerical examples.

\subsection{Generalization of the Framework}
\label{subsec:Gen}
{The objective
~\eqref{eq:OptObj} 
is equal
  to the weighted $\ell_0$ norm of $\bz$:}
\begin{align}
\label{eq:ObjL0}
\sum_{i\in \NN_2}c_i|z_i|_0
\end{align}
where for any real number $z$, $|z|_0=0$ if $z=0$ and $|z|_0=1$
otherwise.  Hence \ref{eq:Opt} admits an equivalent formulation with
the objective function changed to~\eqref{eq:ObjL0} and
with~\eqref{eq:OptCon6} relaxed to $z_i\in[0,1]$, $i\in \NN_2$.

\ref{eq:Opt} can {also} be generalized to
utility functions {of the form:
\begin{align}
  \sum_{i\in \NN_2}c_i|z_i|_0 + f(\br,\bx,\by,\bz)&\label{eq:Opt1Obj}
\end{align}
subject to the same constraints,}
where
$z_i{\in[0,1]}$
indicates the fraction of the total bandwidth used by BTS $i$.
The objective~\eqref{eq:Opt1Obj} is highly versatile.
For example, letting
$
f(\br,\bx,\by,\bz)=\sum_{i\in \NN}b_iz_i
$
accounts for the transmit power consumption being
proportional to the bandwidth allocation;
letting
\begin{equation}
\label{eq:Cost}
f(\br,\bx,\by,\bz)=\beta\sum_{j\in \KK}\frac{\lambda_j}{\sum_{l\in \KK}\lambda_l}\cdot \frac{1}{r_j-\lambda_j},
\end{equation}
incorporates the cost of delay in the objective, where $\beta$ accounts for the tradeoff between energy and delay.

\section{
Reweighted $\ell_1$ Minimization}
\label{sec:Alg}
{With the weighted $\ell_0$ norm~\eqref{eq:ObjL0} as its objective},
\ref{eq:Opt} is a mixed integer program.
{It is generally} difficult to solve due to its combinatorial
nature.
{Such} optimization problems 
{frequently appear} in sparse signal recovery, portfolio
optimization, and statistical estimation.
In this paper, we present an algorithm based on
a low-complexity method, called {\em reweighted $\ell_1$
minimization}
~\cite{candes2008enhancing}.

\subsection{Algorithm Based on Reweighted $\ell_1$ Minimization}

The basic algorithm for solving~\ref{eq:Opt} consists of iterating between solving a convex optimization problem with weighted $\ell_1$ norm relaxation of the $\ell_0$ objective~\eqref{eq:ObjL0}, and updating the weights. The continuous convex optimization problem based on $\ell_1$ relaxation is:
\begin{varsubequations}{P4}
\label{eq:Opt3}
\begin{align}
{\minimize \limits_{\br,\bx,\by,\bz}} \quad& \sum_{i\in \NN_2}w_ic_iz_i &\label{eq:Opt3Obj}\\
\subjectto\;\;
& z_i \ge 0, \qquad i\in \NN_2 \\
&\eqref{eq:OptCon1}\text{--}\eqref{eq:OptCon4} \nonumber
\end{align}
\end{varsubequations}%
where the objective function
{becomes a weighted $\ell_1$ norm of $\bz$ in lieu of the weighted
  $\ell_0$ norm.}

\begin{algorithm}
\caption{Reweighted $\ell_1$ {minimization}.} 
\label{alg:l1}
\begin{algorithmic}[]
\INPUT{$\left(\lambda_j\right)_{j\in \KK}$, $\left(s^{i\to j}_A\right)_{i\in \NN,\;j\in \KK,\;A\subset \NN}$, and $\left(c_i\right)_{i\in \NN_2}$.}
\OUTPUT{$\left(\br,\;\bx,\;\by,\;\bz\right)$.}
\Init{$w_i^1\leftarrow 1, ~\forall i\in \NN_2$,\\  $ u^{-1}\leftarrow0,\; u^0\leftarrow\sum_{i\in \NN_2}c_i,$ and  $t\leftarrow1$.}
\While{$t\leq T$ and $| u^{t-1}-u^{t-2}| > \epsilon_1$}
    \State 1. Compute $\left(\br^{t},\;\bx^{t},\;\by^{t},\;\bz^{t}\right)$ and the corresponding optimum $u^{t}$ by solving~\ref{eq:Opt3} with $\left(w_i^t\right)_{i\in \NN_2}$ as the weights.
    \State 2. Update the weights by $w_i^{t+1}\leftarrow\frac{1}{z_i^{t}+\epsilon_2},~\forall i\in \NN_2$.
    \State 3. $t\leftarrow t+1$.
\EndWhile
\end{algorithmic}
\end{algorithm}

Algorithm~\ref{alg:l1} states the iterative procedure. It starts with the weights $\bw$ as a vector of all
ones. In the $t$th iteration, the algorithm first computes
$\left(\br^{t},\;\bx^{t},\;\by^{t},\;\bz^{t}\right)$ by solving~\eqref{eq:Opt3} with $\bw^t$ as the weights. In fact, this is
a simple linear program.
The weights are {then} updated 
{as} $w_i^{t+1}\leftarrow
1/(z_i^{t}+\epsilon_2)
$, where $\epsilon_2$ is some small number. If $\epsilon_2=0$, the
weight $w_i$ is
{the inverse of} $z_i$.
{Hence} $w_iz_i$
{is a good approximation of} $|z_i|_0$.
We can also view $w_i$ as a penalty term.
If $z_i^{t}$ is large, then $w_i^{t+1} \approx 0$, whereas if $z_i^{t}$ is small, then $w_i^{t+1} \gg 1$, so that $z_i$ is more likely to be driven to 0 in the following iteration. The algorithm terminates if either the maximum number of iterations $T$ is reached or convergence~(according to some predefined threshold $\epsilon_1$) is achieved.\footnote{Here, convergence refers to the objective not necessarily the variables.}

An alternative explanation of Algorithm~\ref{alg:l1} based on the majorization-minimization approach is given in~\cite{pollakis2012base}. It makes use of the following property of the $\ell_0$ norm~\cite{candes2008enhancing,sriperumbudur2011majorization}:
\begin{align}
\label{eq:logL0}
|z|_0=\lim_{\epsilon\to0} g_\epsilon(z)
\end{align}
for $z\ge0$, where
\begin{align}\label{eq:f}
  g_\epsilon(z) = \frac{\log(1+z\epsilon^{-1})}{\log(1+\epsilon^{-1})}.
\end{align}
The majorization-minimization method can be regarded as solving a
sequence of minimization problems, where in each instance it minimizes a
surrogate function that locally majorizes the true objective function (see~\cite{hunter2004tutorial,sriperumbudur2011majorization} for more details). For a concave and differentiable function $f$, a simple and effective majorization-minimization update is:
\begin{align}
\label{eq:MM}
\bz^{t+1}=\arg\min_{\bz\in\zeta}f(\bz^t)+\Delta f(\bz^t)^T(\bz-\bz^t),
\end{align}
where $\zeta$ is the feasible set of the problem. Since
{for every $\epsilon>0$}, $g_\epsilon$
is concave {on $[0,\infty)$},
substituting~\eqref{eq:ObjL0} and~\eqref{eq:logL0}
into~\eqref{eq:MM}, we obtain:
\begin{align}
\label{eq:MML0}
\bz^{t+1}=\arg\min_{\bz\in\zeta}\sum_{i\in \NN_2}\frac{1}{z_i+\epsilon_2}c_iz_i,
\end{align}
where $\zeta$ is the feasible set
{of~\ref{eq:Opt3}}.
It is easy to see that~\eqref{eq:MML0} is exactly what
Algorithm~\ref{alg:l1} computes in each iteration. Hence
Algorithm~\ref{alg:l1} can be viewed as using the
  majorization-minimization method to minimize the approximate
  objective:
{$f(\bz) = \sum_{i\in \NN_2} c_i g_{\epsilon_2}(z_i)$}.
The concavity of $f(\bz)$ and~\eqref{eq:MML0} imply $f(\bz^{t+1})\leq f(\bz^t)$, which establishes the convergence of Algorithm~\ref{alg:l1}.

\subsection{A Refined Algorithm}
\label{subsec:refine}
Algorithm~\ref{alg:l1} mainly deals with the combinatorial nature
of~\ref{eq:Opt} introduced by the binary variables $\bz$. However, the
number of variables in~\ref{eq:Opt3} is $O(kn2^n)$ due to the $2^n$
frequency patterns. We next reduce the complexity and improve the
convergence speed {somewhat}
using the fact that switching each BTS off halves the number of available patterns.

There are two important properties of~\ref{eq:Opt3} and
Algorithm~\ref{alg:l1}. First, if $z_i=0$, according
to~\eqref{eq:OptCon3} and~\eqref{eq:OptCon7}, then $y_A=0$
{for every $A$ that contains $i$}.
Second, if $\epsilon_2$ is close to 0, once Algorithm 1 decides to
switch a certain BTS off in some iteration, it is unlikely to be
reactivated in later iterations. This is because the weight
$
{\epsilon_2}^{-1}$
{corresponding to $z_i=0$ is typically} much larger than
{that corresponding to}
 any other nonzero {element}.
Based on these two properties, a refined version of Algorithm~\ref{alg:l1} is proposed in Algorithm~\ref{alg:refine}, which reduces the dimensions of the feasible set during the iterations.

\begin{algorithm}
\caption{A refined algorithm.}
\label{alg:refine}
\begin{algorithmic}[]
\INPUT{$\left(\lambda_j\right)_{j\in \KK}$, $\left(s^{i\to j}_A\right)_{i\in \NN,\;j\in \KK,\;A\subset \NN}$, and $\left(c_i\right)_{i\in \NN_2}$.}
\OUTPUT{$\left(\br,\;\bx,\;\by,\;\bz\right)$.}
\Init{$w_i^1\leftarrow1,~\forall i\in \NN_2$,\\ $u^{-1}\leftarrow0,\; u^0\leftarrow\sum_{i\in \NN_2}c_i, $ and  $t\leftarrow1$.}
\While{$t\leq T$ and $|u^{t-1}-u^{t-2}| > \epsilon_1$}
    \State 1. Compute $\left(\br^{t},\;\bx^{t},\;\by^{t},\;\bz^{t}\right)$ and the corresponding optimum $u^{t}$ by solving~\ref{eq:Opt3} with $\left(w_i^t\right)_{i\in \NN_2}$ as the weights.
    \State 2. Update the weights by $w_i^{t+1}\leftarrow\frac{1}{z_i^{t}+\epsilon_2}$.
    \If{$\exists i\in \NN_2~\text{s.t. }z_i^{t}=0$, and $\sum_{i:\;z_i^{t}>0}w_i^{t+1}<\frac{\alpha}{\epsilon_2}$}
        \State $\NN_2\leftarrow\{i~|~z_i^{t}>0,\;i\in \NN_2\}$
        \State $\NN\leftarrow \NN_1\cup \NN_2$
    \EndIf
    \State 3. $t\leftarrow t+1$.
\EndWhile
\end{algorithmic}
\end{algorithm}

The only change from Algorithm~\ref{alg:l1} is that the BTSs turned
off at the end of each iteration are eliminated from future
optimizations if a certain condition is met. Namely, if some $z_i$
becomes zero in any iteration, we then compare the sum of the
penalties on all nonzero components, $\sum_{i:z_i>0}w_i^{t+1}$ against
the penalty on any zero component up to a scale factor, namely,
$\alpha/\epsilon_2$.
If the total penalty on the nonzero terms is smaller, the pico BTSs
with zero $z_i$ will be ignored in future iterations, i.e., future
optimizations of~\ref{eq:Opt3} will be considered for the reduced
pico BTS set $\NN'_2=\{i~|~z_i>0,\;i\in \NN_2\}$. Analogous to
Algorithm~\ref{alg:l1}, Algorithm~\ref{alg:refine} can be regarded as
updating a sequence of objective functions $f^t$ in the same form,
but with $\NN_2$ updated in each iteration. With slight abuse of notation,\footnote{$\bz^t$ corresponds to the (possibly reduced) set $\NN_2$. Here $f^1(\bz^t)$ is evaluated by setting any $z_i^t$ not in the original $\NN_2$ to zero.} we have $f^{t}(\bz^{t})=f^1(\bz^{t})$. It is also easy to show that $f^{t+1}(\bz^{t+1}) \leq f^{t+1}(\bz^{t})=f^t(\bz^{t}) \leq f^t(\bz^{t-1})$. Hence the values of the approximate objective $f^1$ after each iteration form a monotonically decreasing sequence, which establishes the convergence of Algorithm~\ref{alg:refine}.

To avoid a premature reduction of the feasible set, the weights on
active and inactive BTSs are compared in
Algorithm~\ref{alg:refine}. When $\sum_{i:z_i^{t+1}>0}w_i^{t+1}<
{\alpha}/{\epsilon_2}$, it is easy to see that an off BTS in the current iteration will not be turned back on in the next iteration according to Algorithm~\ref{alg:l1}. This is because the total reduction of the objective achieved by turning off all currently active BTSs in the next iteration is not enough to compensate for the increased cost due to reactivating a single currently inactive BTS. By setting the scale factor $\alpha$ small enough, the off BTSs that are eliminated from the feasible set are unlikely (albeit still possible) to be turned on again according to Algorithm~\ref{alg:l1}. If $l$ pico BTSs are removed from $\NN_2$, the number of variables is reduced by a factor of $2^l$, which results in a $2^l$ complexity reduction in future iterations. The performance of Algorithms~\ref{alg:l1} and~\ref{alg:refine} will be compared in Section~\ref{sec:Sim}. Algorithms~\ref{alg:l1} and~\ref{alg:refine} can be used to solve~\ref{eq:Opt3} if the function $f$ in~\eqref{eq:Opt3Obj} is convex. The only difference then is that the continuous optimization problem in each iteration is a convex optimization problem instead of a linear program.

\subsection{Post Processing}
\label{subsec:Post}
The post processing in~\cite{pollakis2012base} rounds up the
continuous relaxation of 
{user} associations to binary associations. This is not an issue in
the model considered here since 
{user} association over a slow timescale is indirectly determined
by the amount of resources allocated to each group. 
However, a different post processing can be used to improve the delay performance
without increasing the energy cost, since the
objective~\eqref{eq:OptObj} only depends on the binary variables
$\bz$.

Here post processing is executed after a solution of~\ref{eq:Opt} is
obtained.
The process is performed over the {subset of} active BTSs, {with
  all off BTSs removed from the formulation.}
Specifically,
the post processing is to solve the following 
problem (cf.~\cite{zhuang2015traffic-driven}):
\begin{varsubequations}{P5}
\label{eq:Opt4}
\begin{align}
{\minimize \limits_{\br,\bx,\by}} \quad& \sum_{j\in \KK}\frac{\lambda_j}{\sum_{l\in \KK}\lambda_l}\cdot \frac{1}{r_j-\lambda_j} &\label{eq:Opt4Obj}\\
\subjectto\;\; &\eqref{eq:OptCon1}\text{--}\eqref{eq:OptCon4} \nonumber
\end{align}
\end{varsubequations}%
where objective~\eqref{eq:Opt4Obj} is the average packet sojourn time
in the network.
\ref{eq:Opt4} is a convex {program and is much easier to solve than~\ref{eq:Opt}.}

We shall see in Section~\ref{sec:Sim} that post processing can greatly improve the delay performance in the light traffic regime. The post processing is helpful if the objective only depends on the $\ell_0$ norm.
We can also consider the generalizations in Section~\ref{subsec:Gen}
to minimize the total combined cost of energy and delay.

\section{Numerical Results}
\label{sec:Sim}

In this section, we demonstrate the effectiveness of the proposed
method through numerical examples. In particular,
we compare the performance of Algorithms~\ref{alg:l1} and~\ref{alg:refine} for solving~\ref{eq:Opt} as well as the performance of Algorithm~\ref{alg:l1} for solving~\ref{eq:Opt2}, which 
basically corresponds to the scheme of
~\cite{pollakis2012base}.

\subsection{Simulation Setup}
\label{subsec:setup}
The simulation is carried out over the network in Fig.~\ref{fig:topo}. The HetNet is deployed on a $500 \times 1000\;\text{m}^2$ area. The entire geographic area is divided into a hexagonal grid with $66$ hexagons. Each hexagon represents a user group. The UEs within each hexagon/user group are assumed to be at the center of the hexagon. The BTSs are assumed to be at the vertices of the hexagons. There are two macro BTSs in the HetNet denoted by the dark squares.
Ten pico BTSs are randomly placed in the network denoted by the triangles.

\begin{table}[!t]
\caption{Parameter configurations.}
\label{tab:par}
\centering
\begin{tabular}{||c|c||}
\hline\hline
Parameter & Value/Function\\
\hline
macro transmit power & 46 dBm\\
pico transmit power & 30 dBm\\
total bandwidth & 10 MHz\\
average packet length & 0.5 Mb\\
macro to UE pathloss & $128.1 + 37.6\log_{10}(R)$\\
pico to UE pathloss & $140.7 + 36.7\log_{10}(R)$\\
\hline\hline
\end{tabular}
\end{table}

The spectral efficiency is calculated by~\eqref{eq:SE} with a 30 dB cap on the receive SINR (i.e., an SINR greater than 30 dB is
regarded as 
30 dB). The pathloss models used for macro and pico BTSs are the urban macro (UMa) and urban micro (UMi) models specified in~\cite{3GPP_TR-36.814}. Other parameters used in the simulation are given in Table~\ref{tab:par}, which are also compliant with the LTE standard~\cite{3GPP_TR-36.814}. In the simulation, we only consider pathloss without slow and fast fading.

\subsection{Energy-Efficient Spectrum Allocation}
\label{subsec:Energy}

\begin{figure}[!t]
\includegraphics[width=\mywidth]{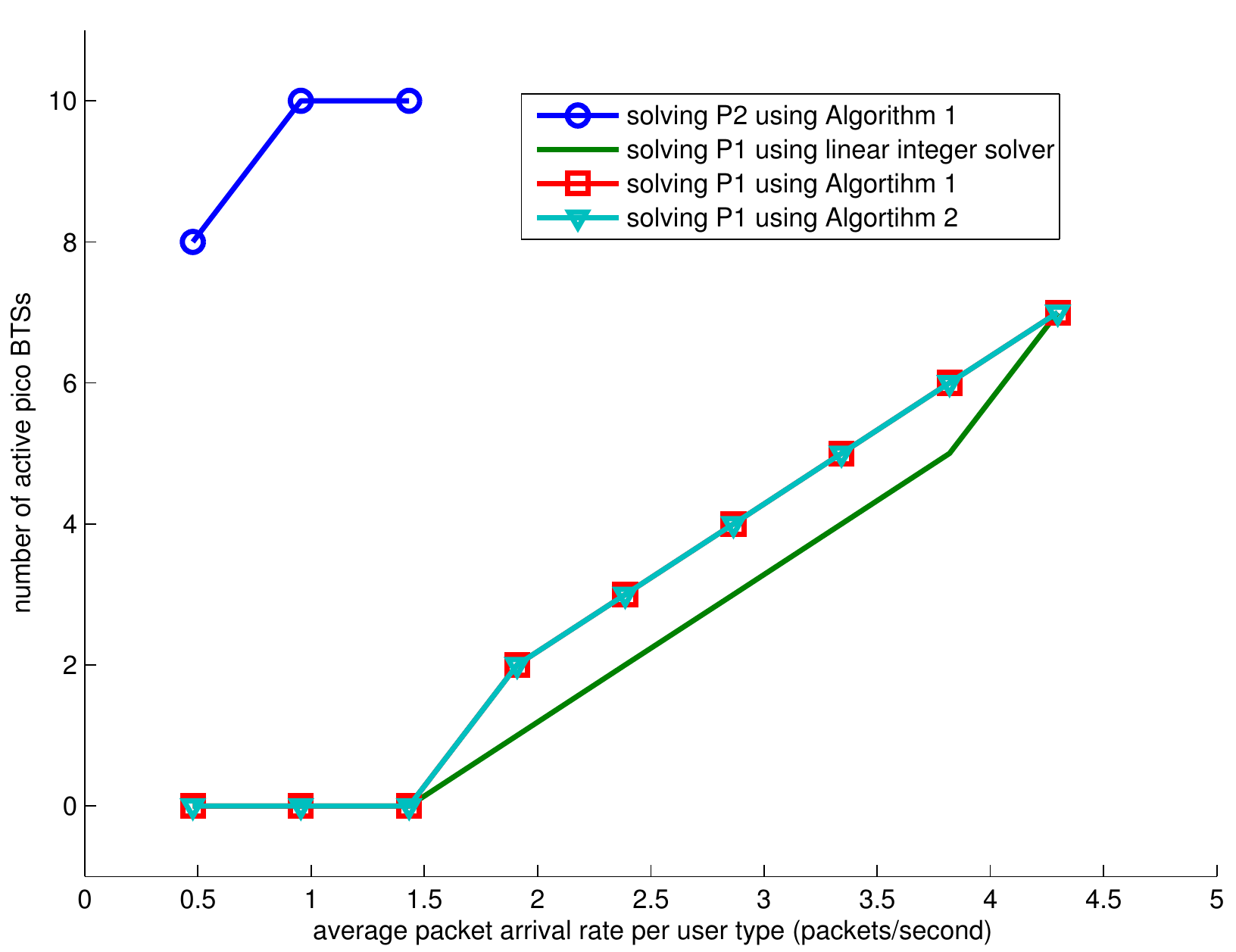}
\centering
\caption{Comparison of energy costs {of different allocation schemes}.
}
\label{fig:EnergyTraffic}
\end{figure}

Fig.~\ref{fig:EnergyTraffic} {illustrates the energy cost due to different allocation schemes at different traffic intensities}.
In the simulation, a random vector $[a_1,\cdots,a_k]$ is first generated with $\mathbb{E}[a_i]=1,~i=1,\cdots,k$. Given the average arrival rate $\bar{\lambda}$, let $\blambda=\bar{\lambda}[a_1,\cdots,a_k]$, i.e., the arrival rates of all user groups are scaled proportionally with $\bar{\lambda}$. The delay requirement for all UEs is 0.5 seconds, i.e., $\tau_j=0.5,\;\forall j\in \KK$. The energy costs of all pico BTSs are set at one unit, i.e., $c_i=1,\;\forall i\in \NN_2$. Hence the $y$ axis also indicates the number of active pico BTSs.
The curve marked by circles is obtained by solving~\ref{eq:Opt2} using
Algorithm~\ref{alg:l1}, which can be interpreted as applying the
scheme 
of~\cite{pollakis2012base} to the scenarios presented in this
paper. The {two} curves marked by squares and triangles are achieved by
solving~\ref{eq:Opt} using Algorithms~\ref{alg:l1}
and~\ref{alg:refine}, respectively.
{These two curves coincide because those}
two algorithms {yield}
the same
energy savings {in this case}.
The curve without any marker is achieved by solving~\ref{eq:Opt} using
a standard integer program solver (to be specified shortly).

According to Fig.~\ref{fig:EnergyTraffic}, the solution
to~\ref{eq:Opt} greatly outperforms the solution to~\ref{eq:Opt2}. The
solution to~\ref{eq:Opt2} can only support up to 1.4 packets/second
per user group. In contrast, the proposed scheme 
{here} can serve up to 4.3 packets/second per user group, a
three-fold throughput gain.

The results obtained using Algorithms~\ref{alg:l1} and~\ref{alg:refine} to solve~\ref{eq:Opt} are very close to the solution obtained using a standard integer program solver. (We observe that at most one extra pico BTS is turned on in all traffic regimes.) Similar results are observed with random realizations of the network and traffic distribution.

\begin{figure*}
\centering
\subfloat[average packet arrival rate = 0.5 packets/second]{
\includegraphics[width =\mywidth]{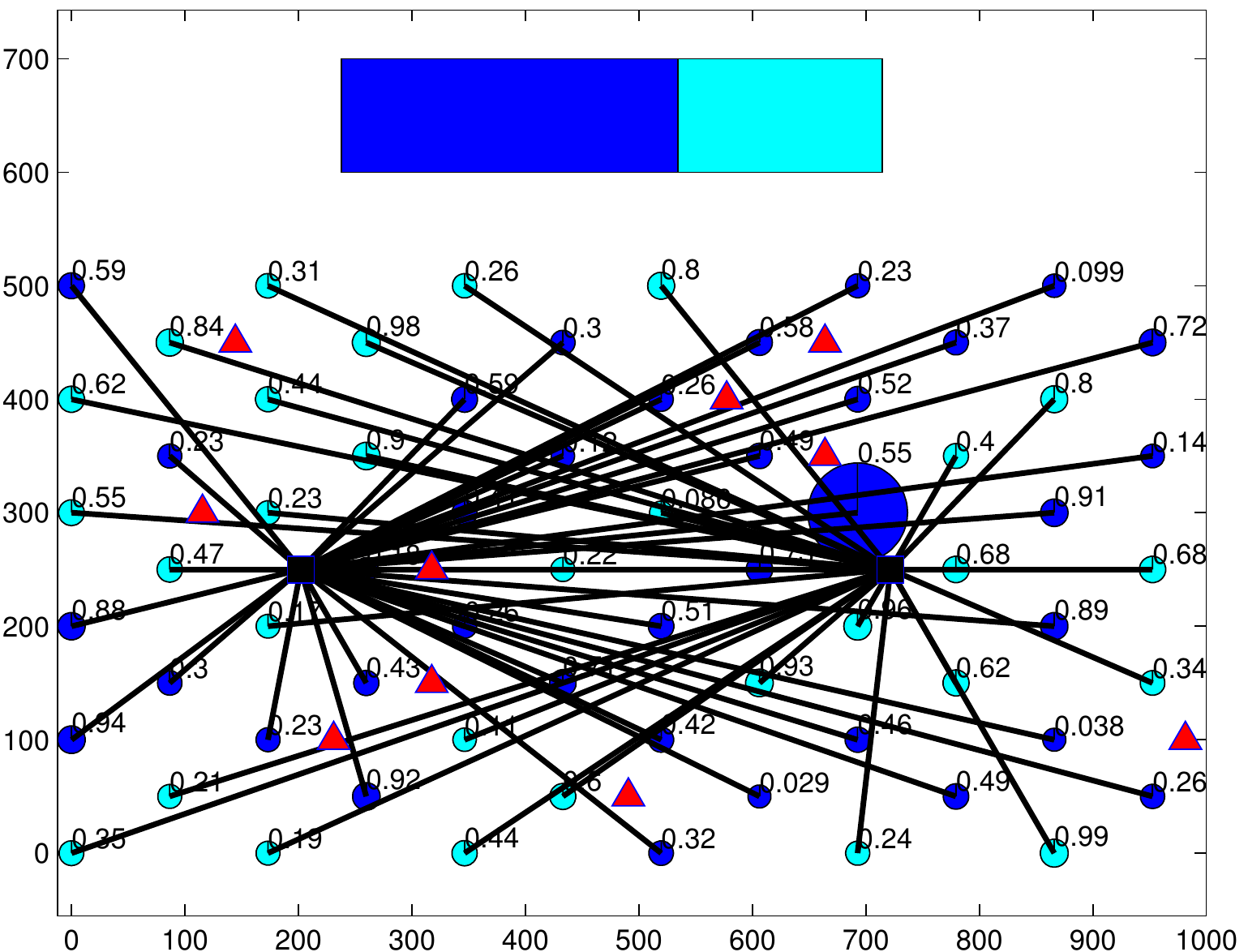}
\label{fig:SpecLow}}
\subfloat[average packet arrival rate = 4.3 packets/second]{
\includegraphics[width =\mywidth]{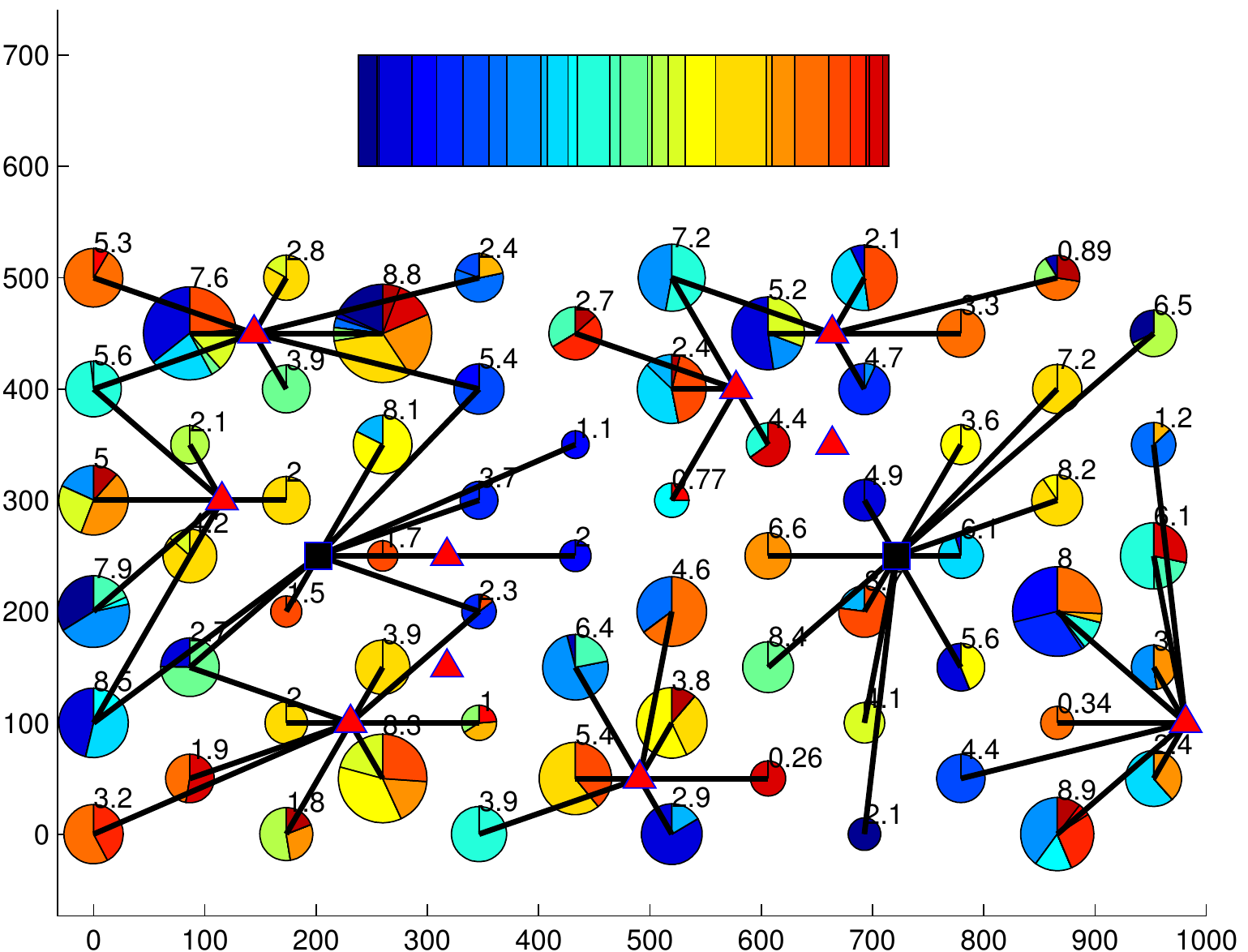}
\label{fig:SpecHigh}}
\caption{Spectrum allocation and user association before post processing in light and heavy traffic regimes.}
\label{fig:SpecAloc}
\end{figure*}

\begin{figure*}
\centering
\subfloat[average packet arrival rate = 0.5 packets/second]{
\includegraphics[width =\mywidth]{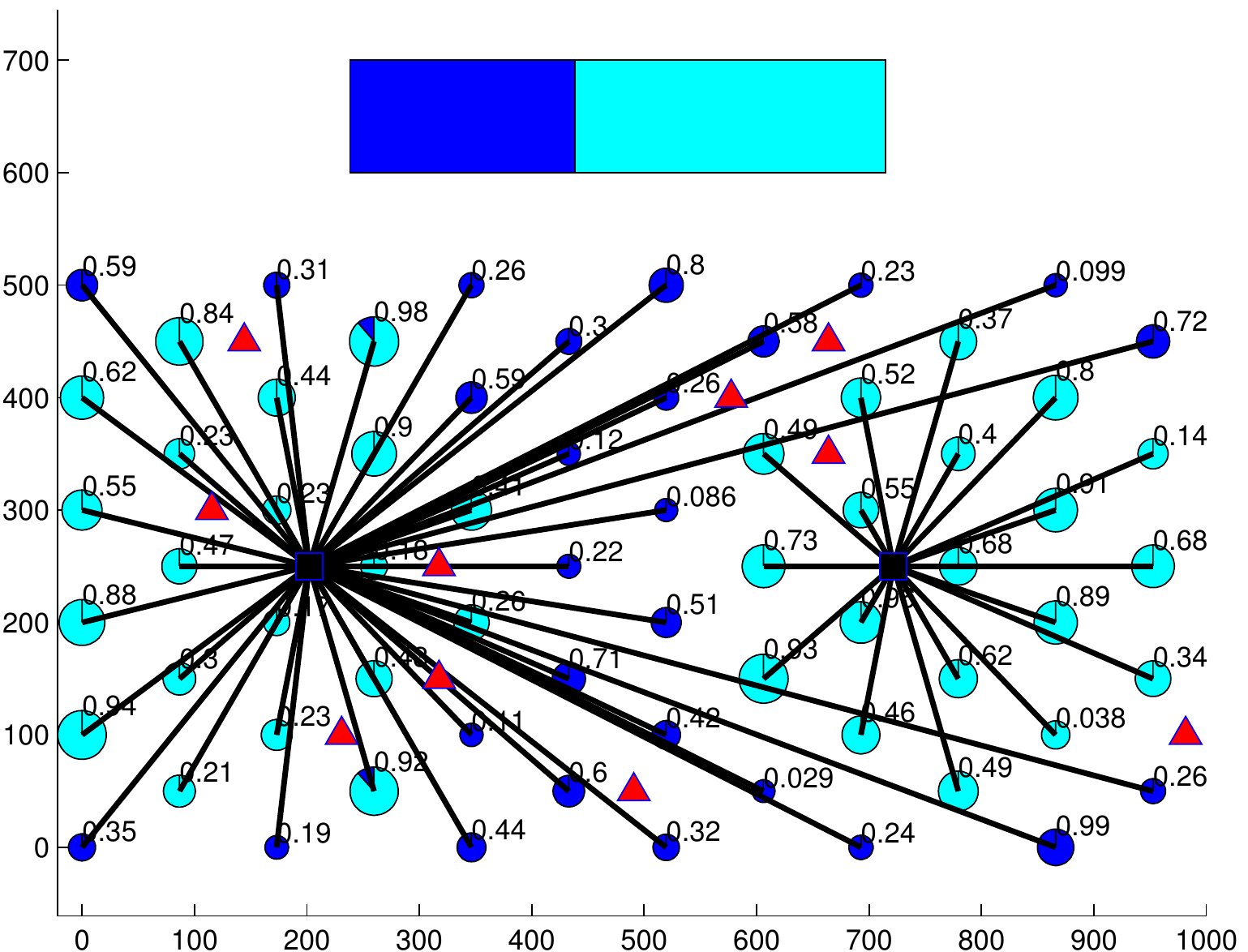}
\label{fig:PostLow}}
\subfloat[average packet arrival rate = 4.3 packets/second]{
\includegraphics[width =\mywidth]{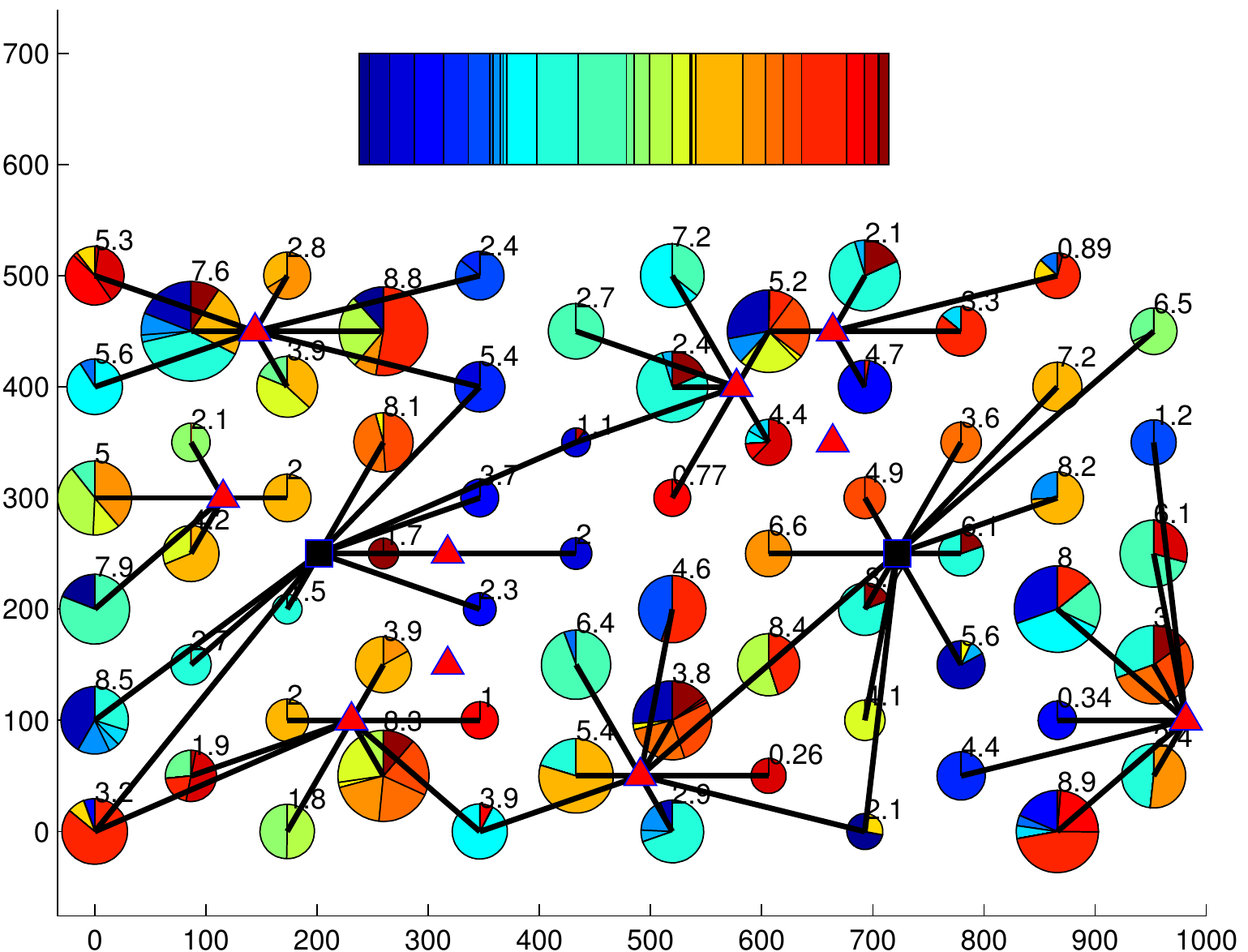}
\label{fig:PostHigh}}
\caption{Spectrum allocation and user association after post processing in light and heavy traffic regimes.}
\label{fig:PostAloc}
\end{figure*}

The spectrum allocation and 
{user} association according to the solution to~\ref{eq:Opt} are
shown in Fig.~\ref{fig:SpecAloc}. Each pie chart indicates the
spectrum allocation at the corresponding user group. The different
colors represent different patterns, and the amount of spectrum
resources allocated to each 
group under each pattern is denoted by the size of the corresponding
pie. The average packet arrival rate of each 
group is shown by the number above the pie chart. Each line segment
joining a BTS and a 
group means the 
group is served by the BTS. The color bars on top of each figure shows the actual spectrum partition into different patterns.

In both Fig.~\ref{fig:SpecLow} and Fig.~\ref{fig:SpecHigh}, the spectrum resources allocated to each user group is roughly proportional to the corresponding traffic demand. The light traffic scenario is shown in Fig.~\ref{fig:SpecLow}. All the pico BTSs are turned off, leaving the 2 macro BTSs to serve all the user groups. The spectrum in Fig.~\ref{fig:SpecLow} is divided into two segments each exclusively used by one of the macro BTSs. Apparently, the allocation is suboptimal in terms of user association, since both macro BTSs intrude into the other cell to serve some user groups that should obviously be served by the other macro BTS. This is because the objective in~\ref{eq:Opt} is only to minimize the number of active pico BTSs. The algorithm terminates as soon as a feasible spectrum allocation is found that satisfies the user delay requirements using the minimum number of pico BTSs. The heavy traffic scenario is shown in Fig.~\ref{fig:SpecHigh}. The spectrum allocation is topology aware. Each pico BTS serves
nearby user groups. Macro BTSs then serve the user groups in coverage holes of the pico BTSs. The two pico BTSs turned off are near clusters of other pico BTSs. The spectrum is orthogonalized among nearby user groups, and is efficiently reused by user groups that are far away.

The spectrum allocations in the light and heavy traffic regimes after
the post processing for further delay improvement, as described in
Section~\ref{subsec:Post}, are shown in Fig.~\ref{fig:PostAloc}. The
allocation in the light traffic regime is shown in
Fig.~\ref{fig:PostLow}. The spectrum is still divided into 2
segments. However, instead of assigning each segment to each BTS
exclusively, one segment is shared by both BTSs and the other is
exclusively used by the macro BTS on the left. The post processing
achieves fractional frequency reuse, i.e., full frequency reuse among
all user groups in cell centers, whereas user groups at cell edges are
served with the spectrum exclusively allocated to the left macro
BTS. The reason that cell edge users in the right macro cell are also
served by the left macro BTS is due to the relatively small network
size.

The spectrum allocation after post processing shown in Fig.~\ref{fig:PostHigh} is similar to that in Fig.~\ref{fig:SpecHigh}. This is because as the traffic gets close to the maximum throughput, there is little margin to further reduce the delay once the delay requirements are met. The post processing reduces the average packet sojourn time in the network from 0.49 seconds to 0.29 seconds in the light traffic regime. However, it only reduces the delay from 0.5 seconds to 0.45 seconds in the heavy traffic regime.

\subsection{Runtime Considerations}
\label{subsec:Prac}
\begin{figure}[!t]
\centering
\includegraphics[width = \mywidth]{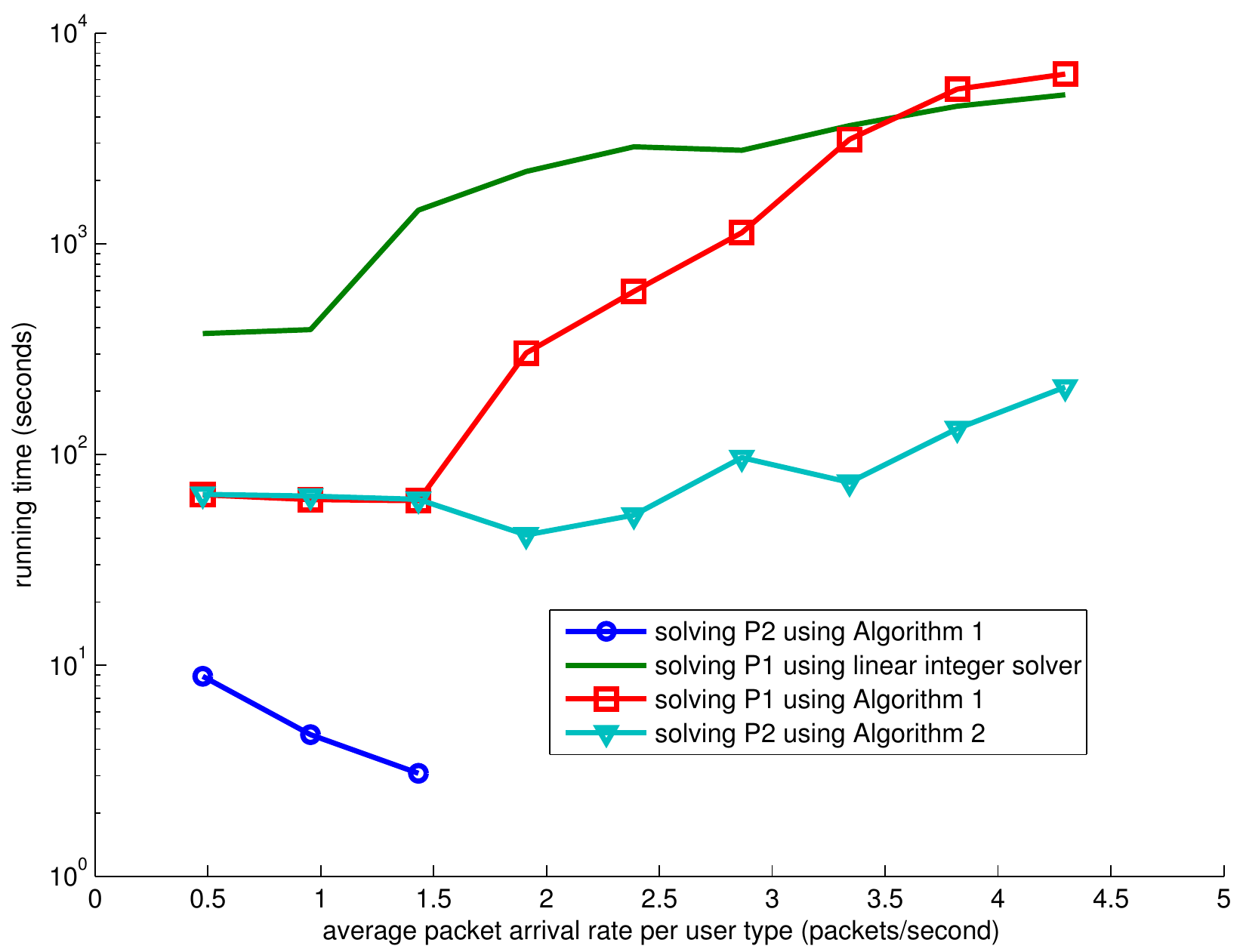}
\caption{Runtime comparison at different traffic intensities.}
\label{fig:time}
\end{figure}

The runtimes of the four schemes are shown in Fig.~\ref{fig:time}. The curves are marked in the same way as in Fig.~\ref{fig:EnergyTraffic}. The optimization problems were solved in Matlab using CVX {from CVX Research, Inc.~on}
an Intel Core i7 2.4 GHz quad-core computer with 16 GB RAM. The continuous linear program~\ref{eq:Opt3} in the iterations of Algorithms~\ref{alg:l1} and~\ref{alg:refine} was solved by the default linear program solver of Gurobi~\cite{gurobi}. The parameters in Algorithms~\ref{alg:l1} and~\ref{alg:refine} are $T=200$, $\epsilon_1=\epsilon_2=10^{-9}$ and $\alpha=0.1$. The curve without marker was solved by the default linear integer program solver of Gurobi.

Although the solution to~\ref{eq:Opt2} is suboptimal in both energy
savings and network throughput, it can be solved much faster than the
other three, and can be applied to large systems with hundreds of
BTSs~\cite{pollakis2012base}.
Among the
{remaining} 
three algorithms,
Algorithm~\ref{alg:refine}'s 
runtime is the most favorable under all traffic conditions.

\begin{table}
\centering
\caption{Number of iterations using Algorithm~\ref{alg:l1} and Algorithm~\ref{alg:refine}.}
\label{tab:iter}
\begin{tabular}{||c||c|c|c|c|c|c|c|c|c||}
\hline\hline
arrival rate (packets/second)& 0.48 & 1.43 & 2.39 &  3.34 & 4.30\\
\hline
iterations for Algorithm~\ref{alg:l1}& 2&  2& 13& 57& 102\\
iterations for Algorithm~\ref{alg:refine}& 2& 2& 9& 6& 13\\
\hline\hline
\end{tabular}
\end{table}

The number of iterations when using Algorithms~\ref{alg:l1}
and~\ref{alg:refine} is shown in Table~\ref{tab:iter}. In the light
traffic regime, both algorithms converge within two iterations. In
fact, both algorithms immediately realize that all pico BTSs should be
turned off under such light loads. Hence the reduction of the feasible
set in Algorithm~\ref{alg:refine} never happens. In the moderate and
heavy traffic regimes, Algorithm~\ref{alg:refine} takes fewer
iterations due to the feasible set reduction. The runtime using
Algorithm~\ref{alg:refine} is reduced by as much as 42 times compared
to Algorithm~\ref{alg:l1} as shown in Fig.~\ref{fig:time}. The shorter
runtime is
{due to both} faster convergence and lower computational complexity in each iteration with reduced dimensions.

\subsection{
{Overhead}}

{A central controller needs to know the traffic intensity of all
UE groups and the spectral efficiency of all links}
(i.e., $\blambda$ and $\bs$)
in order to perform the proposed
energy-efficient global resource management.
Location information 
can help to identify users from different 
groups, which can be acquired using {standard positioning schemes}.
The traffic information for each UE group can be measured at its
serving BTS.
The link gains are routinely measured by the BTSs.  If BTS $i$
  does not receive signals from UE group $j$, then the gain of
  link $i\to j$ can be  
  regarded as zero.

{On the slow timescale considered,}
the overhead of storing and forwarding the aforementioned information over backhaul links is quite small. For example, to describe 10,000 parameters (32 bits each) once every minute translates to about 5 kilobits per second (kbps).
The decision variables need to be fed back to each BTS after
  solving the optimization problem.
The number of variables sent to each BTS is {at most} $O(k^2)$.
{Even with a million variables per minute, the overhead is merely
  500 kbps.}

\section{Conclusion} 
\label{sec:Con}

Traffic-driven radio resource management on a slow timescale has been
proposed and studied for improving energy efficiencies in
HetNets. Joint spectrum allocation, user association, and cell 
activation significantly reduce energy cost and improve system
throughput.
The proposed algorithms can efficiently 
optimize a network
cluster with up to
{20} BTSs.
The improved performance is
at the cost of increased computational complexity for exploiting
spectrum agility.

The main drawback of the current problem formulation 
is that the complexity
  scales exponentially with network size due to the combinatorial
  patterns. In practice, many of the $2^n$ patterns
  in a large HetNet can be easily ruled out, i.e., two BTSs far apart will not
  interfere with each other, and a UE will not be served by a BTS far away.
Extending this approach to {much} larger HetNets 
is {ongoing} work.
The general resource management framework has also been extended to
mixed fast and slow timescales in preliminary work~\cite{teng2015resource}.

\section*{Acknowledgement}
The auhors thank Ermin Wei for pointing out a flaw in an earlier
proof of Theorem 1.

\end{document}